\documentclass{pasj01}

\RequirePackage{mathpazo}
\RequirePackage[T1]{fontenc}

\newcounter{author}
\def\authorcount#1#2{\refstepcounter{author}\label{#1}
                     \altaffiltext{\ref{#1}}{#2}}

\begin{document}
\SetRunningHead{T. Kato et al.}{SU UMa-Type Dwarf Nova with Longest Superhump Period}

\Received{201X/XX/XX}
\Accepted{201X/XX/XX}

\title{OT~J002656.6$+$284933~(CSS101212:002657$+$284933):
      An SU UMa-Type Dwarf Nova with Longest Superhump Period}

\author{Taichi~\textsc{Kato},\altaffilmark{\ref{affil:Kyoto}*}
        Tam\'as~\textsc{Tordai},\altaffilmark{\ref{affil:Polaris}}
        Colin~\textsc{Littlefield},\altaffilmark{\ref{affil:LCO}}
        Kiyoshi~\textsc{Kasai},\altaffilmark{\ref{affil:Kai}}
        Sergey~Yu.~\textsc{Shugarov},\altaffilmark{\ref{affil:Sternberg}}$^,$\altaffilmark{\ref{affil:Slovak}}
        Natalia~\textsc{Katysheva},\altaffilmark{\ref{affil:Sternberg}}
        Anna~M.~\textsc{Zaostrojnykh},\altaffilmark{\ref{affil:Kazan}}
        Roger~D.~\textsc{Pickard},\altaffilmark{\ref{affil:BAAVSS}}$^,$\altaffilmark{\ref{affil:Pickard}}
        Enrique~de~\textsc{Miguel},\altaffilmark{\ref{affil:Miguel}}$^,$\altaffilmark{\ref{affil:Miguel2}}
        Kirill~\textsc{Antonyuk},\altaffilmark{\ref{affil:CrAO}}
        Oksana~\textsc{Antonyuk},\altaffilmark{\ref{affil:CrAO}}
        Elena~P.~\textsc{Pavlenko},\altaffilmark{\ref{affil:CrAO}}
        Nikolai~\textsc{Pit},\altaffilmark{\ref{affil:CrAO}}
        Hiroshi~\textsc{Itoh},\altaffilmark{\ref{affil:Ioh}}
        Javier~\textsc{Ruiz},\altaffilmark{\ref{affil:Ruiz1}}$^,$\altaffilmark{\ref{affil:Ruiz2}}$^,$\altaffilmark{\ref{affil:Ruiz3}}
        Keisuke~\textsc{Isogai},\altaffilmark{\ref{affil:Kyoto}}
        Mariko~\textsc{Kimura},\altaffilmark{\ref{affil:Kyoto}}
        Yasuyuki~\textsc{Wakamatsu},\altaffilmark{\ref{affil:Kyoto}}
        Tonny~\textsc{Vanmunster},\altaffilmark{\ref{affil:Vanmunster}}
        Geoff~\textsc{Stone},\altaffilmark{\ref{affil:AAVSO}}
}

\authorcount{affil:Kyoto}{
     Department of Astronomy, Kyoto University, Kyoto 606-8502, Japan}
\email{$^*$tkato@kusastro.kyoto-u.ac.jp}

\authorcount{affil:Polaris}{
     Polaris Observatory, Hungarian Astronomical Association,
     Laborc utca 2/c, 1037 Budapest, Hungary}

\authorcount{affil:LCO}{
     Department of Physics, University of Notre Dame, 
     225 Nieuwland Science Hall, Notre Dame, Indiana 46556, USA}

\authorcount{affil:Kai}{
     Baselstrasse 133D, CH-4132 Muttenz, Switzerland}

\authorcount{affil:Sternberg}{
     Sternberg Astronomical Institute, Lomonosov Moscow State University, 
     Universitetsky Ave., 13, Moscow 119992, Russia}

\authorcount{affil:Slovak}{
     Astronomical Institute of the Slovak Academy of Sciences,
     05960 Tatranska Lomnica, Slovakia}

\authorcount{affil:Kazan}{
     Institute of Physics, Kazan Federal University,
     Ulitsa Kremlevskaya 16a, Kazan 420008, Russia}

\authorcount{affil:BAAVSS}{
     The British Astronomical Association, Variable Star Section (BAA VSS),
     Burlington House, Piccadilly, London, W1J 0DU, UK}

\authorcount{affil:Pickard}{
     3 The Birches, Shobdon, Leominster, Herefordshire, HR6 9NG, UK}

\authorcount{affil:Miguel}{
     Departamento de Ciencias Integradas, Facultad de Ciencias
     Experimentales, Universidad de Huelva,
     21071 Huelva, Spain}

\authorcount{affil:Miguel2}{
     Center for Backyard Astrophysics, Observatorio del CIECEM,
     Parque Dunar, Matalasca\~nas, 21760 Almonte, Huelva, Spain}

\authorcount{affil:CrAO}{
     Federal State Budget Scientific Institution ``Crimean Astrophysical
     Observatory of RAS'', Nauchny, 298409, Republic of Crimea}

\authorcount{affil:Ioh}{
     Variable Star Observers League in Japan (VSOLJ),
     1001-105 Nishiterakata, Hachioji, Tokyo 192-0153, Japan}

\authorcount{affil:Ruiz1}{
     Observatorio de C\'antabria, Ctra. de Rocamundo s/n, Valderredible, 
     39220 Cantabria, Spain}

\authorcount{affil:Ruiz2}{
     Instituto de F\'{\i}sica de Cantabria (CSIC-UC), Avenida Los Castros s/n, 
     E-39005 Santander, Cantabria, Spain}

\authorcount{affil:Ruiz3}{
     Agrupaci\'on Astron\'omica C\'antabria, Apartado 573,
     39080, Santander, Spain}

\authorcount{affil:Vanmunster}{
     Center for Backyard Astrophysics Belgium, Walhostraat 1A,
     B-3401 Landen, Belgium}

\authorcount{affil:AAVSO}{
     American Association of Variable Star Observers, 49 Bay State Rd.,
     Cambridge, MA 02138, USA}



\maketitle

\begin{abstract}
We observed the 2016 outburst of OT J002656.6$+$284933
(CSS101212:002657$+$284933) and found that it has
the longest recorded [0.13225(1)~d in average]
superhumps among SU UMa-type dwarf novae.
The object is the third known SU UMa-type dwarf nova
above the period gap.  The outburst, however, was unlike
ordinary long-period SU UMa-type dwarf novae
in that it showed two post-outburst rebrightenings.
It showed superhump evolution similar to short-period
SU UMa-type dwarf novae.
We could constrain the mass ratio to less than 0.15
(most likely between 0.10 and 0.15) by using superhump
periods in the early and post-superoutburst stages.
These results suggest the possibility that OT J002656.6$+$284933
has an anomalously undermassive secondary and
it should have passed a different evolutionary track
from the standard one.
\end{abstract}

\section{Introduction}

   SU UMa-type dwarf novae are a class of cataclysmic variables
(CVs) which show superhumps during long-lasting outbursts
called superoutbursts [for general information of CVs,
dwarf novae and SU UMa-type dwarf novae and superhumps,
see e.g. \citet{war95book}].
These superhumps and superoutbursts are now considered
to be a consequence of the 3:1 resonance between
the rotation in the accretion disk and the secondary star
(\cite{whi88tidal}; \cite{osa89suuma}; \cite{hir93SHperiod};
\cite{lub92SH}).  Such a resonance can occur when
the mass-ratio ($q=M_2/M_1$) of the binary is small
enough to accommodate a large accretion disk.
It had long been known that SU UMa-type dwarf novae
are restricted to objects below the famous CV period gap
[cf. \citet{kni06CVsecondary} and \citet{kni11CVdonor}
for the period gap
and modern summary of CV evolution].  The only well-established
exception was, and has long been, one of the earliest known
SU UMa-type dwarf novae, 
TU Men (\cite{sto81tumen1}; \cite{sto84tumen}),
whose orbital period ($P_{\rm orb}$) and superhump period
($P_{\rm SH}$) are 0.1172~d and 0.1257~d, respectively
\citep{men95tumen}.  The lack of SU UMa-type
dwarf novae above the period gap impressed many researchers
and led \citet{whi88tidal} to propose his idea of
tidal instability and its stability condition.
Although more than 700 SU UMa-type dwarf novae have
been identified at the time of writing (cf. \cite{Pdot8}),
TU Men, together with the recently reported faint object
OGLE-GD-DN-009 [$P_{\rm SH}$=0.1310(3)~d, \cite{mro13OGLEDN2}],
have been the only objects above the period gap.
Here, we report on the discovery of an SU UMa-type
dwarf nova which has the longest $P_{\rm SH}$.

   OT J002656.6$+$284933 (hereafter OT J002656)
was discovered as a possible dwarf nova
by the Catalina Real-time Transient Survey
(CRTS; \cite{CRTS})\footnote{
   $<$http://nesssi.cacr.caltech.edu/catalina/$>$.
   For the information of the individual Catalina CVs, see
   $<$http://nesssi.cacr.caltech.edu/catalina/AllCV.html$>$.
} on 2010 December 12 at an unfiltered CCD magnitude
of 18.52 with the detection name CSS101212:002657$+$284933.\footnote{
$<$http://nesssi.cacr.caltech.edu/catalina/20101212/1012121290034112135.html$>$.
}
Due to the faintness, this object did not receive much
attention at the time of this discovery.
There was a faint, blue ($g$=21.6, $u-g$=$-$0.2) SDSS
counterpart SDSS J002656.59$+$284932.9 \citep{SDSS9}
and a GALEX counterpart with a near ultraviolet (NUV)
magnitude of 21.5(3) \citep{GALEX}.
\citet{kat12DNSDSS} estimated the orbital period
to be 0.165(13)~d from SDSS colors using a neural network.
The object was also recorded in outburst at $r$=15.65
on 2002 July 22 by the Carlsberg Meridian Telescope \citep{CMC15}.
The object was reported to be in a bright ($V$=15.5)
outburst on 2013 July 6 by the ASAS-SN \citep{ASASSN} team
(vsnet-alert 15926).  This outburst, however, did not
receive special attention.  The dwarf nova-type classification
has become certain after these multiple outburst detections.

\begin{figure}
  \begin{center}
    \FigureFile(80mm,60mm){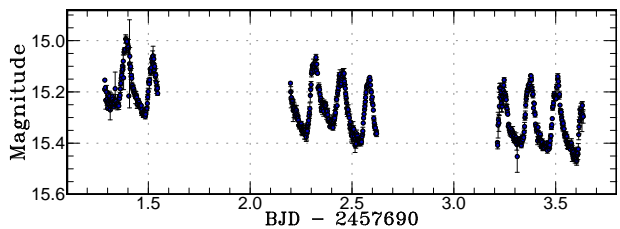}
  \end{center}
  \caption{Superhumps in OT J002656.  The data were
  binned to 0.002~d.}
  \label{fig:j0026shsamp}
\end{figure}

   The object received attention by the detection
of another bright outburst ($V$=14.95)
on 2016 October 23 by the ASAS-SN CV patrol
\citep{dav15ASASSNCVAAS}\footnote{
   $<$http://cv.asassn.astronomy.ohio-state.edu/$>$.
}.  The past light curve in the ASAS-SN CV patrol\footnote{
   The light curve can be seen from
$<$http://cv.asassn.astronomy.ohio-state.edu/outbursts$>$
by searching using the name CSS101212:002657$+$284933.
The CRTS light curve is available at
$<$http://nesssi.cacr.caltech.edu/catalina/20101212/1012121290034112135p.html$>$,
which did not record the bright outburst in 2013.
}
strongly suggested that the 2013 July outburst bore
characteristics of a superoutburst (vsnet-alert 20258)
and an observational campaign of the 2016 outburst
was launched.
Subsequent observations detected 
superhumps (vsnet-alert 20265),\footnote{
   This vsnet-alert announcement was based on the detection
   of a single superhumps by T. Tordai.  Later it became
   evident that K. Kasai had already reported two
   superhump maxima.
}
which further developed into our familiar superhumps
with an astonishingly long ($\sim$0.13~d;
figure \ref{fig:j0026shsamp}; E-figure 1)
$P_{\rm SH}$ (vsnet-alert 20271).
Further observations confirmed the long-period
nature of this object (vsnet-alert 20279,
20286, 20310, 20326, 20331).
The resultant period was longer than that of TU Men.

\section{Observation and Analysis}

   The observations were carried out as in many
campaigns (e.g. \cite{Pdot8})
led by the VSNET Collaboration \citep{VSNET}.
The observers used 20--60cm telescopes during
the outburst and a 1.25-m telescope (Nauchny, Crimea)
after the superoutburst (November 30 and December 5).
All observers used unfiltered CCD cameras except
one run in the $V$-band on November 4
(E-table 1).
They used aperture photometry and extracted magnitudes
relative to comparison stars whose constancy has been 
confirmed by comparison with check stars.  The remaining
small zero-point differences between observers
were corrected by adding constants
to minimize the scatter.
The analysis of superhumps was performed
in the same way as described in \citet{Pdot} and \citet{Pdot6}.
We mainly used R software\footnote{
   The R Foundation for Statistical Computing:\\
   $<$http://cran.r-project.org/$>$.
} for data analysis.
In de-trending the data, we divided the data into
three segments in relation to the outburst phase and
used locally-weighted polynomial regression 
(LOWESS: \cite{LOWESS}, using a smoothing parameter
$f$=0.1) for the superoutburst plateau
and the post-superoutburst period.
For a rapidly fading short segment, we used a linear
fit to remove the trend. 
The times of superhumps maxima were determined by
the template fitting method as described in \citet{Pdot}.
The times of all observations are expressed in 
barycentric Julian days (BJD).

\section{Results}

\subsection{Course of Outburst}

   As shown in the upper panel of figure \ref{fig:j0026humpall},
the object showed the superoutburst plateau until BJD 2457703
(2016 November 10) and then started fading quickly.
The superoutburst lasted for at least 18~d.
Although there was a gap in the observation for the four
subsequent nights due to the interference by the bright
Moon close to this object, we are confident that the object
should have faded further since the object was observed
to be fading quickly on the final night (BJD 2457705)
of the main superoutburst.
The object underwent a post-superoutburst rebrightening
on BJD 2457710 (November 17; E-figure 2) and smoothly faded.
Quite astonishingly, the object had yet another
rebrightening on BJD 2457718 (November 25).
Between these rebrightenings, the object remained
brighter than 19.5 mag, which is 2 mag brighter than
in quiescence (as inferred from the SDSS magnitudes).
Following the second rebrightening, the object became
very faint (fainter than magnitude 20, not plotted
in the figure) and it was likely that the object
returned to quiescence.

\subsection{Superhumps}

\begin{figure}
  \begin{center}
    \FigureFile(80mm,110mm){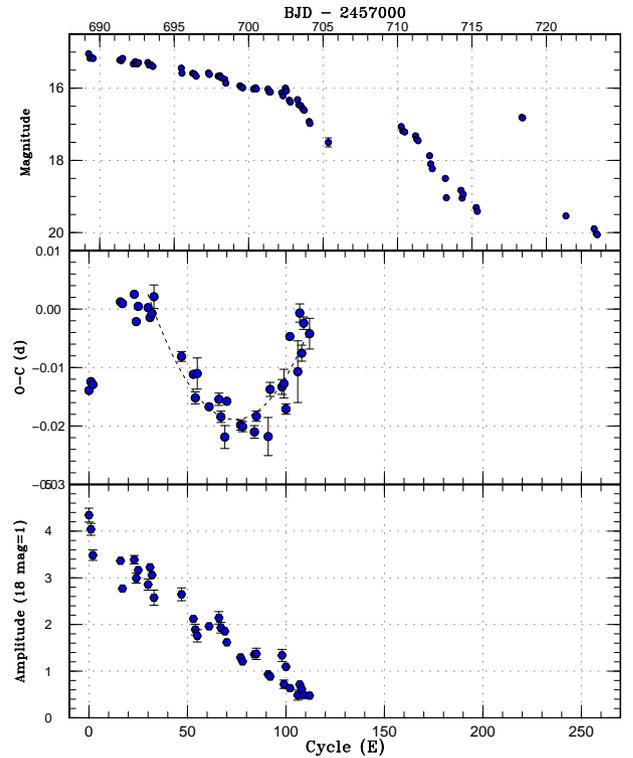}
  \end{center}
  \caption{Light curve and $O-C$ diagram of superhumps in OT J002656.
     (Upper:) Light curve.  The data were binned to 0.1323~d.
     (Middle:) $O-C$ diagram of superhumps.
     We used a period of 0.1323~d for calculating the $O-C$ residuals.
     The dashed curve indicates a parabolic fit for stage B
     superhump (see text).
     (Lower:) Amplitudes of superhumps.  The scale is linear
     and the pulsed flux is shown in a unit corresponding
     to 18 mag = 1.
  }
  \label{fig:j0026humpall}
\end{figure}

   The times of superhump maxima during the superoutburst
plateau are listed in E-table 2.
The $O-C$ diagram (middle panel of figure \ref{fig:j0026humpall})
shows our familiar pattern of stages A and B, characterized
by a rising slope (stage A) and a quadratic curvature
(stage B) as illustrated in figure 4 in \citet{Pdot}
for short-$P_{\rm SH}$ SU UMa-type dwarf novae.\footnote{
   In short-period SU UMa-type dwarf novae, the stage A-B
   transition usually coincides with the peak of superhumps
   amplitudes \citep{Pdot}.  It has become evident that
   in some systems, particularly in long-period systems,
   superhump amplitudes become maxima before the stage A-B
   transition (cf. \cite{kat16v1006cyg}).  We relied on
   the $O-C$ diagram to identify these stages.
}
Stage C (segment with a shorter period following stage B)
was absent or it occurred in the observational gap.
It was, however, certain that stage B to C transition
did not occur before the end of the superoutburst
plateau.  Using the segment before BJD 2457691.6,
which corresponds to stage A in the $O-C$ diagram,
we obtained a superhump period of 0.13320(3)~d
by the phase dispersion minimization (PDM) method \citep{PDM}
(E-figure 3).

   By using the segment 30$\le E \le$112 (which is
well approximated by a parabola in the $O-C$ diagram),
we obtained a positive period derivative of
$P_{\rm dot} = \dot{P}/P$ = $+$16.4(1.6) $\times$ 10$^{-5}$.
The mean superhump period in this segment was 0.13225(1)~d
(E-figure 4).

   Since stage A was not very well observed
and the resultant period may have already been affected
by stage B superhumps,
we independently estimated the period of stage A superhumps
by using an empirical relation that the period of
stage A superhumps is 1.0--1.5\% longer than that of
averaged stage B superhumps \citep{Pdot}
[This relation has been confirmed in \citet{kat13qfromstageA}
for well-observed systems].
This method gives a period of 0.13356--0.13422~d.
Since the lower limit is close to our direct estimate,
we used the upper limit as our upper-limit estimate
of the period of stage A superhumps.

   The superhump signal persisted after the first
rebrightening (see E-table 3; E-figure 2).
We used the segment before BJD 2457716.
The superhump period after BJD 2457710 by the PDM method
was 0.13192(7)~d (E-figure 5).  When we restricted the analysis
after BJD 2457711 and 2457712, we obtained periods
of 0.13183(6)~d and 0.13174(6)~d, respectively.
These values do not greatly differ from each other,
and we adopted a period of 0.1318(1)~d for the period
after the initial rebrightening.

\section{Discussion}

\subsection{Mass Ratio from Superhump Periods}\label{sec:qfromSH}

   As demonstrated in \citet{kat13qfromstageA}, we can
estimate the mass ratio from the fractional superhump
excess of stage A superhumps against the orbital period.
In the case of OT J002656, the orbital period is unknown.
In such a case, we can constrain the mass ratio
by using the periods of stage A superhumps and
post-superoutburst superhumps.

   The dynamical precession rate, $\omega_{\rm dyn}$
in the disk can be expressed by (cf. \cite{hir90SHexcess}):
\begin{equation}
\label{equ:precession}
\omega_{\rm dyn}/\omega_{\rm orb} = Q(q) R(r),
\end{equation}
where $\omega_{\rm orb}$ and $r$ are the angular orbital frequency
and the dimensionless radius measured in units of the binary 
separation $a$.  The dependence on $q$ and $r$ are
\begin{equation}
\label{equ:qpart}
Q(q) = \frac{1}{2}\frac{q}{\sqrt{1+q}},
\end{equation}
and
\begin{equation}
\label{equ:rpart}
R(r) = \frac{1}{2}\sqrt{r} b_{3/2}^{(1)}(r),
\end{equation}
where
$\frac{1}{2}b_{s/2}^{(j)}$ is the Laplace coefficient
\begin{equation}
\label{equ:laplace}
\frac{1}{2}b_{s/2}^{(j)}(r)=\frac{1}{2\pi}\int_0^{2\pi}\frac{\cos(j\phi)d\phi}
{(1+r^2-2r\cos\phi)^{s/2}}.
\end{equation}
This $\omega_{\rm dyn}/\omega_{\rm orb}$ is equal to
the fractional superhump excess in frequency:
$\epsilon^* \equiv 1-P_{\rm orb}/P_{\rm SH}$.

   Following the treatment in \citet{kat13j1222}, we can describe:
\begin{equation}
\label{equ:epsstagea}
\epsilon^*({\rm stage A}) = Q(q) R(r_{\rm 3:1})
\end{equation}
and
\begin{equation}
\label{equ:epspost}
\epsilon^*({\rm post}) = Q(q) R(r_{\rm post}),
\end{equation}
where $r_{\rm 3:1}$ is the radius of the 3:1 resonance
\begin{equation}
\label{equ:radius31}
r_{3:1}=3^{(-2/3)}(1+q)^{-1/3},
\end{equation} 
$\epsilon^*({\rm post})$ and $r_{\rm post}$ are the fractional
superhump excess and disk radius soon after the
the superoutburst, respectively.  By solving equations
(\ref{equ:epsstagea}) and (\ref{equ:epspost}) simultaneously,
we can obtain the relation between $r_{\rm post}$ and $q$.
If we have knowledge about $r_{\rm post}$,
we have a more stringent constraint.

\begin{figure}
  \begin{center}
    \FigureFile(80mm,70mm){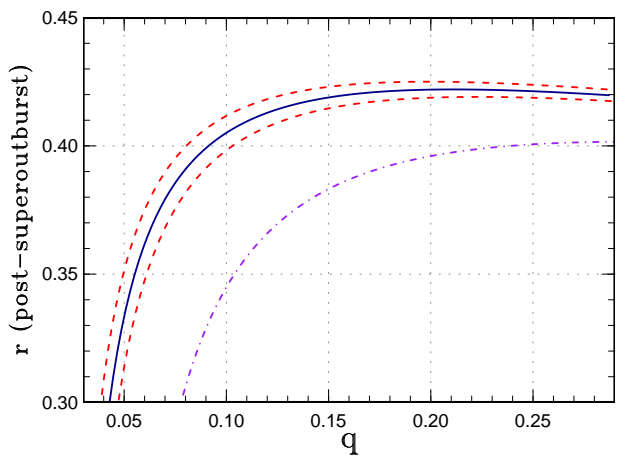}
  \end{center}
  \caption{Relation between $q$ and $r_{\rm post}$ using
  the periods of stage A superhumps and post-superoutburst
  superhumps.  Dashed curves represent the range of
  1$\sigma$ errors.  The long dashed short dashed curve
  corresponds to the upper limit of the period of stage A
  superhumps.}
  \label{fig:qrpost}
\end{figure}

   The result is shown in figure \ref{fig:qrpost}.
The measurements of $r_{\rm post}$ in SU UMa-type
dwarf novae using the same method are within the range of
0.30 and 0.38 \citep{kat13qfromstageA}.
The smaller values represent the values for
WZ Sge-type dwarf novae with multiple rebrightenings
(after such rebrightenings).  It is highly unlikely
that $r_{\rm post}$ is larger than 0.38 in OT J002656,
and it is expected to be somewhat smaller than 0.38
since the object experienced a rebrightening at
the time of our measurement, although it was not after
the final rebrightening.  Figure \ref{fig:qrpost}
indicates $q \le$0.08 for $r_{\rm post}$ = 0.38
and $q \sim$0.06 for $r_{\rm post}$ = 0.34 (this value
was selected as an intermediate radius between
objects without rebrightenings and with multiple
rebrightenings).
Using the upper limit of the period of stage A
superhumps, these values become 0.15 and 0.10,
respectively.
The estimated $P_{\rm orb}$ for $r_{\rm post}$=0.34
is 0.1305~d and 0.1295~d for the period of stage A
superhumps we measured and the upper limit estimated
from stage B superhumps, respectively.
We could not detect a periodic signal from
observations near quiescence (19.9 mag on November 30
and 20.3 mag on December 5, E-figure 6).

   Just for completeness, the $q$ value estimated
from $P_{\rm dot}$ of stage B superhumps using
equation (6) in \citet{kat15wzsge} is 0.13(1)
(the errors reflects only the error of $P_{\rm dot}$), 
although it is not certain whether this equation 
still holds in such a long $P_{\rm orb}$ system.

\subsection{Similarity with WZ Sge-Type Dwarf Novae}

   In addition to the relatively small $q$ suggested
from analysis of superhump periods (subsection \ref{sec:qfromSH}),
the presence of two rebrightenings is also common to
WZ Sge-type dwarf novae \citep{kat15wzsge}.
Very few systems with long $P_{\rm orb}$ show
rebrightenings (cf. \citet{kat16v1006cyg}) and
only two systems V1006 Cyg and OGLE-GD-DN-014
are known to show two rebrightenings \citep{mro13OGLEDN2}.
The $O-C$ diagram of superhumps is also similar
to borderline WZ Sge/SU UMa-type systems in
two respects: (1) absence of stage B-C transition
before the termination of the superoutburst plateau,
(2) relatively large $P_{\rm dot}$ (cf. \cite{kat15wzsge}).

   Although the absence of early superhumps during
the 2016 outburst probably does not favor
the WZ Sge-type classification,
it was not completely excluded since there was
a 4~d gap before the start of our observation
and the phase of early superhumps may have been missed.
Further observations during the next occasion
are encouraged.

\subsection{Evolutionary Status}

   With the estimated $P_{\rm orb} \sim$0.130~d, OT J002656
should have a mass of the secondary of 0.20$M_{\odot}$
if it is on the standard CV evolutionary track
(cf. \cite{kni11CVdonor}), which corresponds to $q \sim$0.25
for an 0.8$M_{\odot}$ white dwarf.
Our observation suggests an unusually low $q$ ($\lesssim$0.15).
We have three possibilities:
(1) the object is a period bouncer, (2) the white dwarf
is exceptionally massive, (3) the object
evolved through an evolutionary path differently from
ordinary CVs.  The possibility (1) appears to be
excluded since the object shows outbursts with
relatively short intervals (at least as short as
3~yr), which cannot be expected for a period bouncer
with a very low mass-transfer rate
(cf. \cite{nak14j0754j2304}).  The relatively short
evolutionary time of superhumps and large amplitude
of superhumps are also signatures disqualifying
a period bouncer (cf. \cite{kat15wzsge}).
Although an extremely massive white dwarf cannot be
excluded, well-determined masses of white dwarfs in
SU UMa-type dwarf novae are in a very narrow region
\citep{sav11CVeclmass} and a very massive white dwarf
appears to be rare.

   It has been observationally known that
some of CVs have undermassive secondaries
(e.g. \cite{tho15asassn13cl}).  Evolutionary models
also suggest that such objects can be formed
if mass transfer occurs after the secondary has undergone
significant nuclear evolution (e.g. \cite{pod03amcvn};
\cite{gol15CVevolution}).  OT J002656 may be
such an object.  Future direct observation
of the secondary or abundance studies may test
this interpretation.

\subsection{Implication on SU UMa-Type Dwarf Novae above Period Gap}

   We found the second known (and with the longest period)
SU UMa-type dwarf nova above the period gap.
The object, however, turned out to be a rather unusual one,
probably not on the standard CV evolutionary track.
It looks likely that the 3:1 resonance is very difficult
to achieve above the period gap in dwarf novae on the standard
evolutionary track.  The result is consistent with
the ``superhump success rate'' in \citet{pat05SH}.
Upon the present discovery, we propose that
the mass ratio and evolutionary state of TU Men needs
to be re-examined using stage A superhumps,
although its superoutbursts
are notoriously rare \citep{bat00tumen}.

\section*{Acknowledgments}

This work was supported by the Grant-in-Aid
``Initiative for High-Dimensional Data-Driven Science through Deepening
of Sparse Modeling'' (25120007) 
from the Ministry of Education, Culture, Sports, 
Science and Technology (MEXT) of Japan.
We acknowledge Prof. M. Uemura for discussion about
CV evolution.
This work was also partially supported by
RFBR grant 15-02-06178 (Crimean team, Shugarov and Katysheva) and
Grant VEGA 2/0008/17, APVV-15-0458 (Shugarov),
NSh-9670.2016.2 (Katysheva team).
We thank R. Stubbings for the literature of TU Men.

\section*{Supporting information}

Additional supporting information can be found in the online version
of this article.
Supplementary data is available at PASJ Journal online.

\section*{Note added in proof (2017 March 21)}

   We have been informed that there is a WZ Sge-type
dwarf nova above the period gap (OGLE-BLG-DN-0174,
Mroz et al. 2015).  The object showed a long outburst
with a long rebrightening followed by three short
rebrightenings in 2010.  There was a short outburst
in 2013 August.  Mroz et al. (2015, Acta Astron.,
65, 313) claimed a superhump period of 0.14474(4)~d using the data
between JD 2455380--2455388 (initial part of
the long outburst).  These observations, however,
were not ideally sampled for detecting superhumps
and there were only 43 points for the segment
JD 2455380--2455388.  Although our own analysis
of the same data detected a period of 0.146(1)~d
(with a significant scatter in the phase diagram),
another period of 0.126(1)~d gave an equally acceptable
phase diagram.  We analyzed the later segment
(JD 2455388--2455395, 38 points) of the later half
of the outburst.  The detected candidate periods
were 0.139~d and 0.119~d.  The periods of the second
segment, if they are indeed true signals, were
shorter than those in the first segment by about 5\%.
Such a large decrease of superhump periods during
a superoutburst has not been recorded in any known system
and we consider that these periods may not be
the true superhump periods.  Although the outburst
light curve strongly suggests a WZ Sge-type
dwarf nova, the data were too insufficient to draw
a firm conclusion about the superhump period.
We therefore do not include this object as
confirmed SU UMa-type dwarf novae above the period gap.
We appreciate the help by P. Mroz for providing
the data and results of their period analysis.


\begin{thebibliography}{}

\bibitem[{Ahn} et~al.(2012)]{SDSS9}
  {Ahn}, C.~P., {et~al.}\ 2012, ApJS, 203, 21

\bibitem[{Bateson} et~al.(2000)]{bat00tumen}
  {Bateson}, F., {McIntosh}, R., \& {Stubbings}, R.\ 2000, Publ.\ Variable\
  Stars\ Sect.\ R.\ Astron.\ Soc.\ New Zealand, 24, 48

\bibitem[{Cleveland}(1979)]{LOWESS}
  {Cleveland}, W.~S.\ 1979, J. Amer. Statist. Assoc., 74, 829

\bibitem[{Davis} et~al.(2015)]{dav15ASASSNCVAAS}
  {Davis}, A.~B., {Shappee}, B.~J., {Archer Shappee}, B., \& {ASAS-SN}\ 2015,
  American\ Astron.\ Soc.\ Meeting\ Abstracts, 225, \#344.02

\bibitem[{Drake} et~al.(2009)]{CRTS}
  {Drake}, A.~J., {et~al.}\ 2009, ApJ, 696, 870

\bibitem[{Goliasch}, {Nelson}(2015)]{gol15CVevolution}
  {Goliasch}, J., \& {Nelson}, L.\ 2015, ApJ, 809, 80

\bibitem[{Hirose}, {Osaki}(1990)]{hir90SHexcess}
  {Hirose}, M., \& {Osaki}, Y.\ 1990, PASJ, 42, 135

\bibitem[{Hirose}, {Osaki}(1993)]{hir93SHperiod}
  {Hirose}, M., \& {Osaki}, Y.\ 1993, PASJ, 45, 595

\bibitem[{Kato}(2015)]{kat15wzsge}
  {Kato}, T.\ 2015, PASJ, 67, 108

\bibitem[{Kato} et~al.(2014)]{Pdot6}
  {Kato}, T., {et~al.}\ 2014, PASJ, 66, 90

\bibitem[{Kato} et~al.(2016a)]{Pdot8}
  {Kato}, T., {et~al.}\ 2016a, PASJ, 68, 65

\bibitem[{Kato} et~al.(2009)]{Pdot}
  {Kato}, T., {et~al.}\ 2009, PASJ, 61, S395

\bibitem[{Kato} et~al.(2012)]{kat12DNSDSS}
  {Kato}, T., {Maehara}, H., \& {Uemura}, M.\ 2012, PASJ, 64, 62

\bibitem[{Kato} et~al.(2013)]{kat13j1222}
  {Kato}, T., {Monard}, B., {Hambsch}, F.-J., {Kiyota}, S., \& {Maehara}, H.\
  2013, PASJ, 65, L11

\bibitem[{Kato}, {Osaki}(2013)]{kat13qfromstageA}
  {Kato}, T., \& {Osaki}, Y.\ 2013, PASJ, 65, 115

\bibitem[{Kato} et~al.(2016b)]{kat16v1006cyg}
  {Kato}, T., {et~al.}\ 2016b, PASJ, 68, L4

\bibitem[Kato et~al.(2004)]{VSNET}
  Kato, T., Uemura, M., Ishioka, R., Nogami, D., Kunjaya, C., Baba, H., \&
  Yamaoka, H.\ 2004, PASJ, 56, S1

\bibitem[{Knigge}(2006)]{kni06CVsecondary}
  {Knigge}, C.\ 2006, MNRAS, 373, 484

\bibitem[{Knigge} et~al.(2011)]{kni11CVdonor}
  {Knigge}, C., {Baraffe}, I., \& {Patterson}, J.\ 2011, ApJS, 194, 28

\bibitem[{Lubow}(1992)]{lub92SH}
  {Lubow}, S.~H.\ 1992, ApJ, 401, 317

\bibitem[{Martin} et~al.(2005)]{GALEX}
  {Martin}, D.~C., {et~al.}\ 2005, ApJ, 619, L1

\bibitem[Mennickent(1995)]{men95tumen}
  Mennickent, R.~E.\ 1995, A\&A, 294, 126

\bibitem[{Mroz} et~al.(2013)]{mro13OGLEDN2}
  {Mroz}, P., {et~al.}\ 2013, Acta\ Astron., 63, 135

\bibitem[{Nakata} et~al.(2014)]{nak14j0754j2304}
  {Nakata}, C., {et~al.}\ 2014, PASJ, 66, 116

\bibitem[{Niels Bohr Institute} et~al.(2014)]{CMC15}
  {Niels Bohr Institute}, U.~o.~C., {Institute of Astronomy}, UK, C., \& {Real
  Instituto y Observatorio de La Armada en San Fernando}\ 2014, VizieR\ Online\
  Data\ Catalog, 1327

\bibitem[{Osaki}(1989)]{osa89suuma}
  {Osaki}, Y.\ 1989, PASJ, 41, 1005

\bibitem[{Patterson} et~al.(2005)]{pat05SH}
  {Patterson}, J., {et~al.}\ 2005, PASP, 117, 1204

\bibitem[Podsiadlowski et~al.(2003)]{pod03amcvn}
  Podsiadlowski, Ph., Han, Z., \& Rappaport, S.\ 2003, MNRAS, 340, 1214

\bibitem[{Savoury} et~al.(2011)]{sav11CVeclmass}
  {Savoury}, C.~D.~J., {et~al.}\ 2011, MNRAS, 415, 2025

\bibitem[{Shappee} et~al.(2014)]{ASASSN}
  {Shappee}, B.~J., {et~al.}\ 2014, ApJ, 788, 48

\bibitem[Stellingwerf(1978)]{PDM}
  Stellingwerf, R.~F.\ 1978, ApJ, 224, 953

\bibitem[Stolz, Schoembs(1981)]{sto81tumen1}
  Stolz, B., \& Schoembs, R.\ 1981, IBVS, 2029

\bibitem[{Stolz}, {Schoembs}(1984)]{sto84tumen}
  {Stolz}, B., \& {Schoembs}, R.\ 1984, A\&A, 132, 187

\bibitem[{Thorstensen}(2015)]{tho15asassn13cl}
  {Thorstensen}, J.~R.\ 2015, PASP, 127, 351

\bibitem[Warner(1995)]{war95book}
  Warner, B.\ 1995, Cataclysmic Variable Stars (Cambridge: Cambridge University
  Press)

\bibitem[Whitehurst(1988)]{whi88tidal}
  Whitehurst, R.\ 1988, MNRAS, 232, 35

\end{thebibliography}
\end{document}